\documentclass[aps,pra,twocolumn,showpacs,preprintnumbers,amsmath,amssymb,footinbib]{revtex4-1}
\usepackage{graphicx}
\usepackage{bm}
\usepackage[colorlinks=true, pdfstartview=FitV, linkcolor=red, citecolor=blue, urlcolor=blue]{hyperref}
\newcommand{\beq}{\begin{equation}}
\newcommand{\eeq}{\end{equation}}
\newcommand{\bea}{\begin{eqnarray}}
\newcommand{\eea}{\end{eqnarray}}

\begin{document}

\preprint{BNL-, RBRC-}

\title{Gross-Witten-Wadia transition in a matrix model of deconfinement}
\author{Robert D. Pisarski}
\email{pisarski@bnl.gov}
\affiliation{
Department of Physics, Brookhaven National Laboratory, 
Upton, NY 11973}
\affiliation{RIKEN/BNL, Brookhaven National Laboratory, 
Upton, NY 11973}
\author{Vladimir V. Skokov}
\email{vskokov@quark.phy.bnl.gov}
\affiliation{
Department of Physics, Brookhaven National Laboratory, 
Upton, NY 11973}
\begin{abstract}
We study the deconfining phase transition at nonzero temperature
in a $SU(N)$ gauge theory, using a matrix
model which was analyzed previously at small $N$.
We show that the model is soluble at infinite $N$, and
exhibits a Gross-Witten-Wadia transition.
In some ways, the deconfining phase transition is of first order:
at a temperature $T_d$,
the Polyakov loop jumps discontinuously from $0$ to 
$\frac{1}{2}$, and there is a nonzero
latent heat $\sim N^2$.
In other ways, the transition is of second order: {\it e.g.},
the specific heat diverges as 
$C \sim 1/(T-T_d)^{3/5}$ when $T \rightarrow T_d^+$.
Other critical exponents satisfy the usual scaling relations of a second
order phase transition.
In the presence of 
a nonzero background field $h$ for the Polyakov loop, there is a phase
transition at the temperature $T_h$ where the value of the 
loop $= \frac{1}{2}$, with $T_h < T_d$.  Since
$\partial C/\partial T \sim 1/(T-T_h)^{1/2}$ as
$T \rightarrow T_h^+$, this transition is of third order.
\end{abstract}
\maketitle

The properties of the deconfining phase transition for a $SU(N)$ gauge
theory at nonzero temperature 
are of fundamental interest.  At small $N$, this transition can
only be understood through numerical simulations on the lattice
\cite{DeTar:2009ef, *Petreczky:2012rq}.  Large $N$ can be studied through
numerical simulations
\cite{Lucini:2005vg, *Teper:2008yi, *Panero:2009tv, *Datta:2010sq, *Mykkanen:2011kz, *Lucini:2012wq} and in reduced models
\cite{Makeenko:2004bz, *Narayanan:2003fc, *Kiskis:2003rd, *Narayanan:2007dv, *GonzalezArroyo:2010ss, *GonzalezArroyo:2012fx}.
In the pure glue
theory, this transition can be modeled
through an effective model, such as a matrix model
\cite{Pisarski:2000eq, Dumitru:2003hp, *Oswald:2005vr, *Pisarski:2006hz, *Hidaka:2008dr, *Hidaka:2009hs, *Hidaka:2009xh, *Hidaka:2009ma, Meisinger:2001cq, *Meisinger:2001fi, Dumitru:2010mj, Dumitru:2012fw, Kashiwa:2012wa, Sasaki:2012bi, *Ruggieri:2012ny, *Diakonov:2012dx}.

One limit in which the theory can be solved analytically is
by putting it on a sphere of femto-scale dimensions
\cite{Sundborg:1999ue, *Aharony:2003sx, *Aharony:2005bq, *Schnitzer:2004qt, *AlvarezGaume:2005fv, *AlvarezGaume:2006jg, *Hollowood:2009sy, Dumitru:2004gd}.  
An effective theory is constructed directly by integrating out all modes
with nonzero momentum, and gives a matrix model which
is soluble at large $N$ 
\cite{Brezin:1977sv, Gross:1980he, Wadia:1979vk, *Wadia:1980cp, Jurkiewicz:1982iz}.  As a function
of temperature,
it exhibits a Gross-Witten-Wadia transition 
\footnote{The original analysis of 
Gross and Witten \cite{Gross:1980he, Wadia:1979vk, *Wadia:1980cp} considered
a lattice gauge theory in $1+1$ dimensions.  
For the Wilson action with lattice coupling $\beta$, there is a third
order phase transition with respect to 
$\beta$.  Different lattice actions give
a variety of phase transitions, but again with respect
to the lattice couplings \cite{Jurkiewicz:1982iz}.
This is in contrast to the femtosphere (at infinite $N$)
or the present model (at any $N$), 
which are true thermodynamic phase transitions
with respect to temperature.}.  
That is, it exhibits
aspects of both first order {\it and} second order phase
transitions; thus it can be 
termed ``critical first order'' \cite{Dumitru:2004gd}.
Since the theory has finite spatial volume, however, 
there is only a true phase transition
at infinite $N$.  Thus on a femtosphere,
the Gross-Witten-Wadia transition appears to be mere curiosity.

Matrix models have been developed as an effective theory 
for deconfinement in four spacetime dimensions (and infinite volume).
These models, which involve
zero \cite{Meisinger:2001cq, *Meisinger:2001fi}, one
\cite{Dumitru:2010mj}, and two parameters 
\cite{Dumitru:2012fw, Kashiwa:2012wa}, are soluble analytically for
two and three colors, and numerically for four or more colors.
In this paper we show that these models are also soluble analytically
for infinite $N$.  Most unexpectedly, we find that the model
exhibits a Gross-Witten-Wadia transition, very similar
to that on a femtosphere.
This is surprising because on a femtosphere, the matrix model is
dominated by the Vandermonde determinant, and looks nothing like
the matrix models of Refs.
\cite{Meisinger:2001cq, *Meisinger:2001fi, Dumitru:2010mj, Dumitru:2012fw, Kashiwa:2012wa}.  
This suggests that the Gross-Witten-Wadia transition may not be an artifact
of a femtosphere, but might occur for $SU(\infty)$ gauge theories
in infinite volume.
At the end of this paper we estimate how large $N$ must be to see
signs of the Gross-Witten-Wadia transition at infinite $N$.

\section{Zero background field}
\label{sec:critical}

We expand about a constant background field for the vector potential,
\beq
A_0^{i j} = \frac{2 \pi T}{g} \; q_i \; \delta^{i j} \; ,
\label{define_A0}
\eeq
where $i,j = 1\ldots N$.  This $A_0$ field is a diagonal $SU(N)$ matrix,
and so $\sum_{i = 1}^N q_i = 0$.  The thermal Wilson line
is the matrix ${\bf L} = \exp(2 \pi i {\bf q})$; its trace
is the Polyakov loop in the fundamental representation,
$\ell_1 = {\rm tr} \, {\bf L}/N$.
At any $N$, this represents a possible ansatz for the region
where the expectation value of the Polyakov loop is less than unity.
This region has been
termed the ``semi'' quark gluon plasma (QGP)
\cite{Dumitru:2003hp, *Oswald:2005vr, *Pisarski:2006hz, *Hidaka:2008dr, *Hidaka:2009hs, *Hidaka:2009xh, *Hidaka:2009ma}.
At infinite $N$, this ansatz is the simplest possible
for the master field in the semi-QGP.

The potential we take is a sum of two terms,
\beq
\widetilde{V}_{{\rm eff}}(q) \; = \; - \; d_1(T) \; \widetilde{V}_1(q) \; 
+ \; d_2(T) \; \widetilde{V}_2(q) \; ,
\label{eff_potential}
\eeq
where
\beq
\widetilde{V}_n(q) 
= \sum_{i,j = 1}^{N_c} |q_i - q_j|^n (1 - |q_i - q_j|)^n \; .
\label{define_V}
\eeq
The term $\sim \widetilde{V}_2(q)$ 
is generated perturbatively at one loop order; that
$\sim \widetilde{V}_1(q)$ is added to drive the 
transition to the confined phase.  
Previously, the functions $d_1$ and $d_2$ were chosen as
$
d_1(T) = (2 \pi^2/15) \,  c_1 \, T^2 T_d^2
$ and
$
d_2(T) = 2 \pi^2/3 \, ( T^4 - c_2 \, T^2 T_d^2 )
$ ,
where $T_d$ is the temperature for deconfinement
\cite{Dumitru:2010mj, Dumitru:2012fw, Kashiwa:2012wa}.
These matrix models also included
terms independent of the $q$'s, $\sim c_3 \, T^2 \, T_d^2$ 
and $\sim B \, T_d^4$.

The values of these parameters were chosen to agree with 
results from numerical simulations
on the lattice \cite{Dumitru:2010mj, Dumitru:2012fw, Kashiwa:2012wa}.
As we show, however, when $N$ is infinite, at the transition temperature
the nature of
the solution is independent not only of the values of these parameters,
but even of the choice of the functions $d_1(T)$ and $d_2(T)$ (modulo
modest assumptions, given later).

The matrix model in Eqs. (\ref{eff_potential}) and (\ref{define_V}) is
rather different from that on a femtosphere
\cite{Sundborg:1999ue, *Aharony:2003sx, *Aharony:2005bq, *Schnitzer:2004qt, *AlvarezGaume:2005fv, *AlvarezGaume:2006jg, *Hollowood:2009sy, Dumitru:2004gd}.  
On a femtosphere the
dominant term driving confinement is the Vandermonde determinant,
$\sim \Pi_{i,j} \log| \exp(2 \pi i q_i) - \exp(2 \pi i q_j) |$;
in the present model it is the terms $\sim \widetilde{V}_n(q)$.
The logarithmic singularities of the
Vandermonde determinant are stronger than those of the
from the absolute values $\sim |q_i - q_j|^n$ in the $\sim \widetilde{V}_n(q)$.

To treat infinite $N$, we introduce the variable $x = i/N$, so that
$q_i \rightarrow q(x)$, and the potential is
an integral over $x$. 
It is useful to introduce the eigenvalue density, 
$\rho(q) = dx/dq$  \cite{Brezin:1977sv}.  
The integrals over $x$ then become integrals over $q$,
weighted by $\rho(q)$.  The eigenvalue density must be positive, and
by definition is normalized to 
\beq
\int^{q_0}_{-q_0} \; dq \; \rho(q) = 1  \; .
\label{eigenvalue_density}
\eeq
Polyakov loops are traces of powers of the thermal Wilson line,
\beq
\ell_j = \frac{1}{N}\, 
{\rm tr} \, {\bf L}^j =
\int^{q_0}_{-q_0} \; dq \; \rho(q) \; \cos(2 \pi j q) \; .
\label{loopj}
\eeq
As noted before, the first Polyakov loop, $\ell_1$, is that in the fundamental 
representation.  For $j \geq 2$, 
the relationship of the $\ell_j$ to Polyakov loops in irreducible
representations is more involved \cite{Dumitru:2003hp}, 
but all $\ell_j$ are gauge invariant, and so physical quantities.

By a global $O(2)$ rotation we can assume that the expectation value of
$\ell_1$ is real.
Consequently, we take $\rho(q)$ to be even in $q$,
$\rho(q) = \rho(-q)$.  Anticipating the results,
we also assume that the integral over $q$
does not run the full range
from $-\frac{1}{2}$ to $\frac{1}{2}$, 
but only over a limited range, from $-q_0$ to $+q_0$.  

Going to integrals over $q$, 
we can take out overall factors of $N^2$ from the potentials,
with $\widetilde{V}_n(q) = N^2 \, V_n(q)$, where
\beq
V_n(q) = \int dq \int dq' \, \rho(q) \, \rho(q')
|q - q'|^n (1 - |q - q'|)^n  \; .
\label{potential}
\eeq
In this expression and henceforth, all integrals over $q$ run from $-q_0$
to $+ q_0$, as in Eqs. (\ref{eigenvalue_density}) and (\ref{loopj}).

We then define $\widetilde{V}_{{\rm eff}}(q) = N^2 V_{{\rm eff}}(q)$, where
$V_{{\rm eff}} = -d_1 V_1 + d_2 V_2$.  
Solving the model at infinite $N$, then, is just
a matter of finding the (minimal) stationary point
of $V_{{\rm eff}}(q)$ with respect to the $q_i$'s.

The equations of motion follow by differentiating 
the potential in Eq. (\ref{eff_potential}) with respect to
$q_i$, and then taking
the large $N$ limit.  Doing so, we find
$$
0 = \left[d_1 + d_2\right] \, q \, -\frac{d_1}{2} \int dq' 
\rho(q') \, {\rm sign}(q - q')
$$
\beq
+ d_2 \int dq' \rho(q') 
\left[ - 3 (q - q') |q - q'| + 2 (q - q')^3 \right] \; , 
\label{derpot1}
\eeq
where ${\rm sign}(x) = \pm 1$ for $x \gtrless 0$.
For simplicity we write
$d_1(T)$ and $d_2(T)$ just as $d_1$ and $d_2$.

To solve the equation of motion in Eq. (\ref{derpot1}), 
we follow Jurkiewicz and Zalewski \cite{Jurkiewicz:1982iz} and
use the following trick.  What is difficult is that Eq. (\ref{derpot1})
is an integral equation for $\rho(q)$.  To reduce this to
a differential equation, take 
$\partial/\partial q$ of Eq. (\ref{derpot1}),
\beq
0 = d_1 + d_2 - d_1 \, \rho(q) 
\label{derpot2}
\eeq
$$
+ 6 \, d_2 \int dq' \rho(q') 
\left[ - (q - q') {\rm sign}(q - q') + (q - q')^2 \right] \; .
$$
Notice that this does not give us the second variation of the potential
with respect to an arbitrary variation of $q$, which is related
to the mass squared.  Instead, we take the derivative of the equation of
motion, with respect to a solution of the same.

We then continue until we eliminate any integral over $q'$.
Taking $\partial/\partial q$ of Eq. (\ref{derpot2}) gives
\beq
d_1 \, \frac{d\rho(q)}{dq} = 
6 \, d_2 \, \int dq' \rho(q') 
\left[ - {\rm sign}(q - q') + 2 (q - q') \right] \; .
\label{derpot3}
\eeq
Lastly, by taking one final derivative, we obtain
\beq
\frac{d^2}{dq^2}  \rho(q) + d^2 \, [ \rho(q) - 1 ] \; = 0 \; .
\label{derpot4}
\eeq
In this expression we introduce the ratio 
\beq
d^2(T) = \frac{12 \; d_2(T)}{d_1(T)} \; .
\eeq
We assume that like the solution at small $N$
\cite{Dumitru:2010mj, Dumitru:2012fw, Kashiwa:2012wa}, 
that $d(T)$ increases with $T$, and 
$d(T) \rightarrow \infty$ as $T \rightarrow \infty$.
We note that the only detailed property of $d(T)$ which we
require is that its expansion about $T_d$ is
linear in $T - T_d$.  This is a minimal assumption which is standard
in mean field theory.  

We thus need to solve Eqs. (\ref{derpot1}) - (\ref{derpot4}), subject
to the condition of Eq. (\ref{eigenvalue_density}).  The solution
of Eq. (\ref{derpot4}) is trivial, 
\beq
\rho(q) = 1 + b \cos( d \, q) \;\;\; , \;\;\;
q: -q_0 \rightarrow q_0 \; ,
\label{soln_rho}
\eeq
where $b$ is a constant to be determined.  We assume that
$\rho(q) = 0$ for $|q| > q_0$.
We have checked numerically that a multi gap 
solution \cite{Jurkiewicz:1982iz}, where $\rho(q) \neq 0$
over a set of gaps in $q$, does not minimize the potential;
see the discussion at the end of Sec. (\ref{sec:nonh_noncrit}).

When $q_0 < \frac{1}{2}$, $\rho(q_0) \neq 0$, and the solution drops discontinuously
to zero at the endpoints.  This stepwise discontinuity is charactertistic
of the model, and presumably reflects the singularities from the absolute
values in the potential.  

The eigenvalue density 
in Eq. (\ref{soln_rho}) is
simpler than that in the Gross-Witten model 
\cite{Sundborg:1999ue, *Aharony:2003sx, *Aharony:2005bq, *Schnitzer:2004qt, *AlvarezGaume:2005fv, *AlvarezGaume:2006jg, *Hollowood:2009sy, Dumitru:2004gd, Gross:1980he, Wadia:1979vk, *Wadia:1980cp, Jurkiewicz:1982iz}, where
\beq
\rho_{GW}(q) = \frac{1}{2} \cos(\pi q)
\left[ 1 - \frac{\sin^2(\pi q)}{\sin^2(\pi q_0)} \right]^{1/2} \; .
\label{Gross_Witten_density}
\eeq
For any $q_0$, this vanishes at the endpoints, $\rho_{GW}(\pm q_0) = 0$,
while at the transition, $q_0 = \frac{1}{2}$.
Due to the Vandermonde
determinant in the potential,
the density $\rho_{GW}(q)$ has a nontrivial analytic structure
in the complex $q$-plane, while $\rho(q)$ does not.
Since the Vandermonde potential is so different 
from $V_{{\rm eff}}$, though,
it is natural to find that 
$\rho_{GW}(q)$ is unlike $\rho(q)$ in Eq. (\ref{soln_rho}).

Eq. (\ref{soln_rho}) solves Eq. (\ref{derpot3}) without further constraint.
To solve the remaining equations, remember that all integrals run from 
$-q_0 \rightarrow q_0$.  
The normalization condition of Eq. (\ref{eigenvalue_density}) gives
$b \sin(d \, q_0) = d( \frac{1}{2} - q_0 )$.  
After some algebra, one can show that Eqs. 
(\ref{derpot1}) and (\ref{derpot2}) are equivalent, with the solution
\beq
\cot( d \, q_0) = \frac{d}{3} \left( \frac{1}{2} - q_0 \right)
- \frac{1}{d \left(1/2 - q_0 \right) } \; ,
\label{soln_q0}
\eeq
and
\beq
b^2 = \frac{d^4}{9} \left( \frac{1}{2} - q_0 \right)^4
+ \frac{d^2}{3} \left( \frac{1}{2} - q_0 \right)^2 + 1 \; .
\label{soln_b}
\eeq
Thus in the end, we only have to solve
two coupled algebraic equations, Eqs. (\ref{soln_q0}) and (\ref{soln_b}),
for $q_0$ and $b$ as functions of $d = d(T)$.

At low temperature, $d$ is small, and the theory is in the confined
phase, where $b = 0$ and $q_0 = \frac{1}{2}$.
The eigenvalue density is constant, $\rho(q) = 1$,
and all Polyakov loops vanish, $\ell_j = 0$.
Thus the confined phase is characterized by
the maximal repulsion of eigenvalues.  
The Gross-Witten model also has a constant eigenvalue density in
the confined phase, which is expected, as only a constant
eigenvalue density gives $\ell_j = 0$ for all loops.

In the limit of high temperature $d \rightarrow \infty$.  The solution
is $q_0 = 6/d^2$ and $b = d^2/12$.
The eigenvalue density is $\rho \approx d^2/12$, which becomes a
delta-function $\delta(q)$ for infinite $d$.  That is, at high
temperatures all eigenvalues coalesce into the origin, and 
all Polyakov loops equal one, $\ell_j = 1$.  

As the temperature and so $d(T)$ is lowered, the transition occurs
when $q_0 = \frac{1}{2}$, for which $d(T_d) = 2 \pi$.  At the transition
point, the eigenvalue density is
\beq
\rho(q) = 1 + \cos(2 \pi q) \; \;\; ; \;\;\; T = T_d \; .
\label{critical_rho}
\eeq
From Eq. (\ref{loopj}), 
\beq
\ell_1(T_d^+) = \frac{1}{2} \;\;\; , \;\;\;
\ell_j(T_d) = 0 \; , j \geq 2 \; .
\label{loops_confined}
\eeq
Thus at the transition, only 
the Polyakov loop in the fundamental representation is nonzero,
equal to $\frac{1}{2}$.

What is unforeseen is that at $T_d^+$, the eigenvalue density
in the present model, Eq. (\ref{critical_rho}), coincides
{\it identically} with that in the Gross-Witten model, 
Eq. (\ref{Gross_Witten_density}).
Consequently, properties exactly at $T_d^+$, such as the expectation
values of the $\ell_j$, are the same in the two models.  Since they
differ away from $T_d$, other properties are similar, but not
necessarily identical.

Consider the behavior in the deconfined phase just above the 
transition point, taking
$
d = 2 \pi (1 + \delta d)
$.
The solution is $q_0^s = \frac{1}{2} (1 - \delta q)$, where
\beq
\delta q = \left(\frac{45}{\pi^4}\right)^{1/5} \delta d^{1/5}
+ \frac{1}{7} \left(\frac{375}{\pi^2}\right)^{1/5} \delta d^{3/5}
+ \frac{25}{49} \delta d + \ldots \; ,
\label{dq_zeroh}
\eeq
\beq
b = 1 + \frac{1}{2} \left(\frac{25 \pi^2}{3}\right)^{1/5} \delta d^{2/5}
+ \frac{ 29 }{56} 
\left(\frac{25 \pi^2}{3}\right)^{2/5} \, \delta d^{4/5} 
+ \ldots \; .
\label{b_zeroh}
\eeq
Using this, one finds that
\beq
\ell_1 = \frac{1}{2} 
+ \frac{1}{4} \left(\frac{25 \pi^2}{3}\right)^{1/5} \, \delta d^{2/5} 
+ \ldots \; ,
\label{ell1_zeroh}
\eeq
while all $\ell_j \sim \delta d$ for $j \geq 2$.

Remember that at $T_d$,
$\ell_1$ jumps discontinuously, from $0$ to $\frac{1}{2}$, 
as expected for a first order transition.
Assuming that $\delta d \sim T_d - T$, though, Eq. (\ref{ell1_zeroh}) shows
that as $T \rightarrow T_d^+$, 
\beq 
\ell_1(T) - \frac{1}{2} \sim (T_d - T)^\beta \;\;\; , \;\;\;
\beta = \frac{2}{5} \; .
\label{beta}
\eeq
That is, near the transition $\ell_1(T)$ exhibits a 
power like behavior which is characteristic of a second
order phase transition --- although $\ell_1(T_d^+) \neq 0$.

For arbitrary $d$, after some algebra one finds that 
at $q_0^s$, the solution of Eqs. 
(\ref{soln_q0}) and (\ref{soln_b}), the potential equals
\beq
V_{{\rm eff}}(q_0^s)  - V_{{\rm eff}}^{{\rm conf}} =
- \; d_2 \; \frac{16 }{15} \left( \frac{1}{2} - q_0^s \right)^5  \; .
\label{potential_Td}
\eeq
The potential in the confined phase is
$V_{{\rm eff}}^{{\rm conf}} = V_{{\rm eff}}(\frac{1}{2}) = - d_1/6 + d_2/30$.
In these matrix models, the pressure is
\beq
p(T) = - V_{{\rm eff}}(q_0^s) + V_{{\rm eff}}^{{\rm conf}} \; .
\label{pressure}
\eeq
This subtraction ensures that 
the pressure, and the associated energy density,
are suppressed by $\sim 1/N^2$ in the confined phase.
In the models of Ref. \cite{Dumitru:2010mj, Dumitru:2012fw},
$V_{{\rm eff}}^{{\rm conf}}$ is given by the term $\sim c_3$.
Expanding about $T_d$, 
\beq
V_{{\rm eff}}(q_0) - V_{{\rm eff}}^{{\rm conf}} =
- \; \frac{48 d_2}{\pi^4} \, \delta d \; 
- \; \frac{270 d_2}{7 \pi^3}\left( 
\frac{25}{2 \pi^{3}}\right)^{1/5} \delta d^{7/5} + \ldots
\label{expansion_potential_about_Td}
\eeq
Assuming that $\delta d \sim T - T_d$, as is true of
the functions in Refs. \cite{Dumitru:2010mj, Dumitru:2012fw, Kashiwa:2012wa},
the leading term in Eq. (\ref{expansion_potential_about_Td})
$\sim \delta d$ shows that the first
derivative of the pressure with respect to temperature, 
which is related to the energy density $e(T)$,
is nonzero at $T_d^+$.  
Since the pressure and the energy density 
are suppressed by $\sim 1/N^2$ in the confined phase,
the latent heat is nonzero and $\sim N^2$, $\sim e(T_d^+)$.

Using the explicit forms for $d_1(T)$ and $d_2(T)$, 
we find that the latent heat is
$e(T_d^+)/(N^2 T_d^4) = 1/\pi^2 \sim .10..$.
This is about four times 
smaller than the lattice results of Ref. \cite{Datta:2010sq}
who find $\sim 0.39$ for the same quantity.
The lattice results 
can be accomodated by adding a term like a MIT bag constant to the model
\cite{Dumitru:2012fw}.  
Such a term is $\sim T_d^4$ but independent of the $q$'s, and so only
changes the latent heat, but does not affect any other result.

The second term in Eq. (\ref{expansion_potential_about_Td})
shows that the
second derivative of the pressure with respect to temperature 
diverges as $T \rightarrow T_d^+$, 
\beq
\frac{\partial^2 }{\partial T^2} \; p(T)
 \sim \frac{1}{(T - T_d)^\alpha} \; \; \; , \;\;\;
\alpha = \frac{3}{5} \; .
\label{specific_heat}
\eeq
This is the usual divergence of the specific heat for a second
order phase transition.

\section{Nonzero background field, $T = T_d$}
\label{sec:nonh_crit}

Background fields can be added for each loop $\ell_j$.  In this
paper we just consider a background field for the simplest loop,
$\ell_1$, since only that is nonzero at $T_d$, Eq. (\ref{loops_confined}).
We add
\beq
V_h(q) = -\; \frac{d_1}{(2 \pi)^2} \;  h \; \ell_1 \; 
\label{def_backgrd_field}
\eeq
to the potential $V_{{\rm eff}}(q)$, and find the solution as before.
After taking three derivatives of the equation of motion, with
respect to a solution, we obtain the analogy of Eq. (\ref{derpot4}),
\beq
\frac{d^2}{dq^2}  \rho(q) + d^2 \, [ \rho(q) - 1] 
+ (2 \pi)^2 \; h \; \cos(2 \pi q) \; = 0 \; .
\label{eom_nonzeroh}
\eeq
This equation is valid for any $d$.  It is necessary to treat
the case of $T_d$, where $d = 2 \pi$, seperately from $T \neq T_d$.

In this section we consider the point of phase transition, where
$d = 2 \pi$.  The solution of Eq. (\ref{eom_nonzeroh}) is
\beq
\rho(q) = 1 + b \, \cos( 2 \pi \, q) - \pi \, h \; q \; \sin(2 \pi q) \; ,
\label{soln_rho_backh_Td}
\eeq
where $q: -q_0 \rightarrow q_0$.  
Notice that the $h$-dependent term $q \, \sin(2 \pi q)$ arises
because when $T= T_d$, Eq. (\ref{eom_nonzeroh}) represents
a driven oscillator at  the  resonance frequency. The value of the constants $b$ and
$q_0$ now depend upon both $d(T)$ and the background field, $h$.

The analogy of Eq. 
(\ref{derpot3}) is solved by Eq. (\ref{soln_rho_backh_Td}).
The normalization condition, 
Eq. (\ref{eigenvalue_density}), plus the analogy of Eq. (\ref{derpot2}),
gives two equations for $b$ and $q_0$; as before, Eq. (\ref{derpot1}) does not
give a new condition.

When $h \neq 0$,
the explicit form of the analogy of 
Eq. (\ref{eigenvalue_density}) is elementary,
but that of Eq. (\ref{derpot2}) is rather ungainly.  We thus
present the results of the solution in the limit of small background
field, $h \ll 1$.  We find that $q_0^s = \frac{1}{2} (1 - \delta q)$, where
\beq
\delta q = \left(\frac{45}{2 \, \pi^4}\right)^{1/5} \; h^{1/5}
+ \frac{3}{14} 
\left(\frac{3}{200 \, \pi^2} \right)^{1/5} \; h^{3/5}
+ \ldots
\label{deltaq_nonzeroh_Td}
\eeq
and
\beq
b = 1 + \frac{1}{2} \left(\frac{25 \, \pi^2}{12}\right)^{1/5} \; h^{2/5} 
+ \frac{39}{56} \left(\frac{27 \pi^4}{80}\right)^{1/5} \; h^{4/5}
+ \ldots
\label{b_nonzeroh_Td}
\eeq

For this solution, at the minimum the $h$-dependence of the potential is
\beq
V_{{\rm eff}}(q_0^s, h) = 
- \frac{d_1}{8 \pi^2} h
+ \frac{d_1}{112 \pi} \left( \frac{25}{12 \, \pi^3} \right)^{1/5}
h^{7/5} + \ldots \; .
\label{potential_Td_h}
\eeq
The expectation value of the loop $\ell_1$ is
\beq
\ell_1 = \frac{1}{2}
+ \frac{1}{4} \left(\frac{25 \, \pi^2}{12}\right)^{1/5} h^{2/5}
+ \frac{39}{112} \left( \frac{27 \pi^4}{80} \right)^{1/5} h^{4/5}
+ \ldots \; .
\label{ell1_Td_h}
\eeq
Hence $\ell_1 - \frac{1}{2} \sim h^{1/\delta}$, where $\delta = 5/2$.  This
shows that the critical exponents of this model satisfy the usual
Griffths scaling relation,
\beq
2 - \alpha = \beta (1 + \delta) \; .
\label{scaling_relations}
\eeq

The effective potential, as a function of $\ell_1$,
is computed by taking the Legendre transform,
\beq
\Gamma(\ell_1) = V_{{\rm eff}}(h) + \frac{d_1}{(2 \pi)^2} h_1 \ell_1 \; .
\label{eff_potential_define}
\eeq
Expanding the potential in 
$\delta \ell_1 = \ell_1 - \frac{1}{2}$ at $T_d^+$,
\beq
\Gamma(\ell_1) = 
+ \frac{128 \, \sqrt{3} \, d_1}{35 \, \pi^3}
\; \delta \ell_1^{7/2}
+ \frac{32 \, d_1}{5}
\; \delta \ell_1^{4} + \ldots 
\label{eff_pot_Td_onehalf}
\eeq
This is a {\it very} flat potential, starting
only as $(\ell_1 - \frac{1}{2})^{7/2}$.  This is in contrast to the 
femtosphere,
where the potential behaves as $\sim (\ell_1 - \frac{1}{2})^3$ 
about the similar point \cite{Aharony:2003sx,Dumitru:2004gd}.  

Expanding at $T_d^-$ gives the expansion of the potential about $\ell_1 = 0$.
One can show, and we verify in the next section, that this potential
vanishes.  This implies that the potential has an unusual form:
it is zero from $\ell_1: 0 \rightarrow \frac{1}{2}$, and then turns on as
in Eq. (\ref{eff_pot_Td_onehalf}).  
Graphically, this potential is
like that on the femtosphere; see, {\it e.g.}, Fig. (1) of
Ref.~\cite{Dumitru:2004gd}.  

\section{Nonzero background field, $T \neq T_d$}
\label{sec:nonh_noncrit}

Consider now the theory in a nonzero background field for $\ell_1$,
Eq. (\ref{def_backgrd_field}), away from the transition, so $d \neq 2 \pi$.
The eigenvalue density again solves Eq. (\ref{eom_nonzeroh}).  The
solution is simpler when $d \neq 2 \pi$, and is just the sum of the
solution for $h = 0$ and an $h$-dependent term,
\beq
\rho(q) = 1 + b \, \cos(d \, q) + \frac{1}{1 - (d/2 \pi)^2} \; h
\; \cos(2 \pi q) \; .
\label{eigen_dens_away_Td_h}
\eeq
The solution follows as previously, and we simply summarize the results.

We first consider the confined phase, defined to
be the solution for which $q_0 = \frac{1}{2}$ and $b = 0$.
The expectation value of the loop $\ell_1$ is
\beq
\ell_1 = \frac{1}{1 - (d/2 \pi)^2} \; \frac{h}{2} \; .
\label{vev_ell_conf}
\eeq
For this solution the potential equals
\beq
V_{{\rm eff}}^{{\rm conf}}(h) - V_{{\rm eff}}^{{\rm conf}} 
= + \frac{1}{1- (d/2 \pi)^2}
\; \frac{h^2}{8 \pi^2} \; .
\label{potential_h_conf}
\eeq
Performing the Legendre transformation, we find
\beq
\Gamma(\ell_1) = \left( 1 - \frac{d^2}{4 \pi^2} \right) \frac{1}{\pi^2}
\; \ell_1^2 \; .
\label{eff_pot_conf_h}
\eeq
This shows that in the confined phase, when $d < 2 \pi$
the mass squared of the $\ell_1$ loop is positive, as expected.
It also shows that this mass vanishes at $T_d$ when $h = 0$; this
justifies the statements about the potential at the end of 
the previous section.

Consider a special value of $d$, $d_h^2 = 4 \pi^2 (1 - h)$;
the corresponding temperature is defined to be $T_h$,
$d(T_h) = d_h$.  At this 
temperature, the eigenvalue density
of Eq. (\ref{eigen_dens_away_Td_h}) coincides exactly with that
at the transition in zero background field, Eq. (\ref{critical_rho}).
Notably, the values of the loop at $h \neq 0$ and $T = T_h$
are the same as for $h = 0$ and $T = T_d$: $\ell_1(T_h) = \frac{1}{2}$,
with $\ell_j = 0$ for $j \geq 2$, Eq. (\ref{loops_confined}).
Thus we may suspect that something special happens at $T = T_h$.
For example, the confined phase is only an acceptable solution when
$T < T_h$, as only then is the eigenvalue density positive definite.  

This suggests that a phase transition occurs at $d_h$.  To show this,
we compute for about this value of $d$, taking
$d^2 = d_h^2 + 4 \pi^2 \, h \, \delta d$.  Solving the model as before,
in the deconfined phase the solution is $q_0^s = \frac{1}{2} (1 - \delta q)$,
where
\beq
\delta q = \frac{1}{\pi} \left( \frac{3}{2} \right)^{1/2} \delta d^{1/2}
+ \frac{\sqrt{6}}{40 \pi} \; (8 h - 5) \; \delta d^{3/2} + \ldots
\label{deltaq_away_trans_deconf}
\eeq
\beq
b = - \frac{4}{5} \sqrt{6} \; \left( 1 - h \right)^{3/2}
 {\rm csc}(\sqrt{1 - h} \; \pi) \; \delta d^{5/2} + \ldots 
\label{b_away_trans_deconf}
\eeq
With this results we compute the potential in the deconfined phase, to find
\beq 
V_{{\rm eff}}(h) - V_{{\rm eff}}^{{\rm conf}}(h) =
- \frac{3 \sqrt{6}} {5 \, \pi^3} \; \delta d^{5/2} + \ldots
\label{potential_away_trans_h}
\eeq

Taking $\delta d \sim T_h - T$, we find that the {\it third} derivative
of the pressure, with respect to temperature, diverges at $T_h$,
\beq
\frac{\partial^3 }{\partial T^3} \; p(T)
 \sim \frac{1}{(T - T_h)^{1/2}} \; \;\; , \;\;\;
T \rightarrow T_h^+ \; .
\eeq

In zero background field, then, there is a critical first order transition at a
temperature $T_d$.  Turning on a background field $\sim h \, \ell_1$, 
the first order transition is 
immediately wiped out for any $h \neq 0$.  Even so,
there remains a third order phase transition, at a temperature
$T_h < T_d$, where the expectation value of the loop $\ell_1 = \frac{1}{2}$.
This behavior is the same as on a femtosphere
\cite{Aharony:2003sx, Schnitzer:2004qt, Dumitru:2004gd}.  

In principle one can also add a background field for any loop, $\ell_j$
for $j \geq 2$.  It is direct to derive the equations of motion
and obtain a solution for the eigenvalue density.  Obtaining
the minimum of the potential is not elementary, though.
The original model of Gross and Witten \cite{Gross:1980he} 
involves the Vandermonde determinant plus a term $\sim |{\rm tr}{\bf L}|^2$.
The solution for the eigenvalue density is a 
function which is nonzero on one interval, between $-q_0$ and $q_0$.
Jurkiewicz and Zalewski \cite{Jurkiewicz:1982iz} showed that 
when terms such as 
$|{\rm tr}\, {\bf L}^2|^2$ are added to the Gross-Witten model, that
in general it involves
functions which are nonzero on more than one interval.  We have checked
numerically that when only $h_1 \neq 0$, that such multi-gap
solutions do not minimize the potential.  
We do find, however, that multi-gap solutions do minimize the potential
in the presence of background fields for
$\ell_j$ when $j \geq 2$.  
Since only $\ell_1 \neq 0$ at $T_d$ and $T_h$, we defer the
problem of background $\ell_j$ for $j \geq 2$.

\section{Finite N}
\label{sec:finiteN}

The model can be solved numerically at finite $N$.
This confirms, as expected on general grounds \cite{Dumitru:2012fw},
that the deconfining transition is of first order for any $N \geq 3$.
It also shows that the critical behavior found
at infinite $N$ is smoothed out at large but finite $N$.

Using the numerical solution of the model,
in the Figure we show the behavior of the specific heat, divided by
$N^2-1$, for different values of $N$.  
To see the putative divergence of the specific heat at infinite $N$,
rather large values of $N$ are necessary, $N \geq 40$.

This Figure also shows that the increase in the specific heat 
only manifests itself very close to the transition, 
within $\sim 0.2\%$ of $T_d$.  At present, direct numerical simulations 
on the lattice treat moderate values of $N \sim 4-10$ 
\cite{Lucini:2005vg, *Teper:2008yi, *Panero:2009tv, *Datta:2010sq, *Mykkanen:2011kz, *Lucini:2012wq}.   
For most quantities there seems to be a weak variation with $N$.

The present matrix model suggests that {\it very} near $T_d$, 
a novel phase transition may occur at large $N$.  The values of $N$ 
at which critical first order behavior arise can presumably be studied only
in reduced models
\cite{Makeenko:2004bz, *Narayanan:2003fc, *Kiskis:2003rd, *Narayanan:2007dv}.  

This begs the question of whether or not the Gross-Witten-Wadia transition 
does in fact occur at infinite $N$ \footnote{
It is also possible to develop a matrix model to study deconfinement
in three spacetime dimensions \cite{PSS:2012}.  
This involves different functions of $q$ than those
in four dimensions, Eq. (\ref{define_V}).  It is not clear if this
model is soluble analytically at infinite $N$, or if it exhibits a
Gross-Witten-Wadia transition in that limit.}.
On the femtosphere, one can easily solve the model
in the presence of additional couplings, such as 
$(|{\rm tr}\, {\bf L}|^2)^2$.  
Such additional couplings turn the Gross-Witten-Wadia transition into an
ordinary first order transition \cite{Dumitru:2004gd}.  
We have not been able to solve the
present model in the presence of additional couplings.

The most likely possibility is that as on a femtosphere, the
presence of additional couplings washes out the Gross-Witten-Wadia transition.
Nevertheless, gauge theories are remarkable things.  Certainly
it is worth studying $SU(N)$ gauge theories, at very large values
of $N$, to see if there is a Gross-Witten-Wadia transition at infinite $N$.

\begin{figure}
\includegraphics[width=0.47\textwidth]{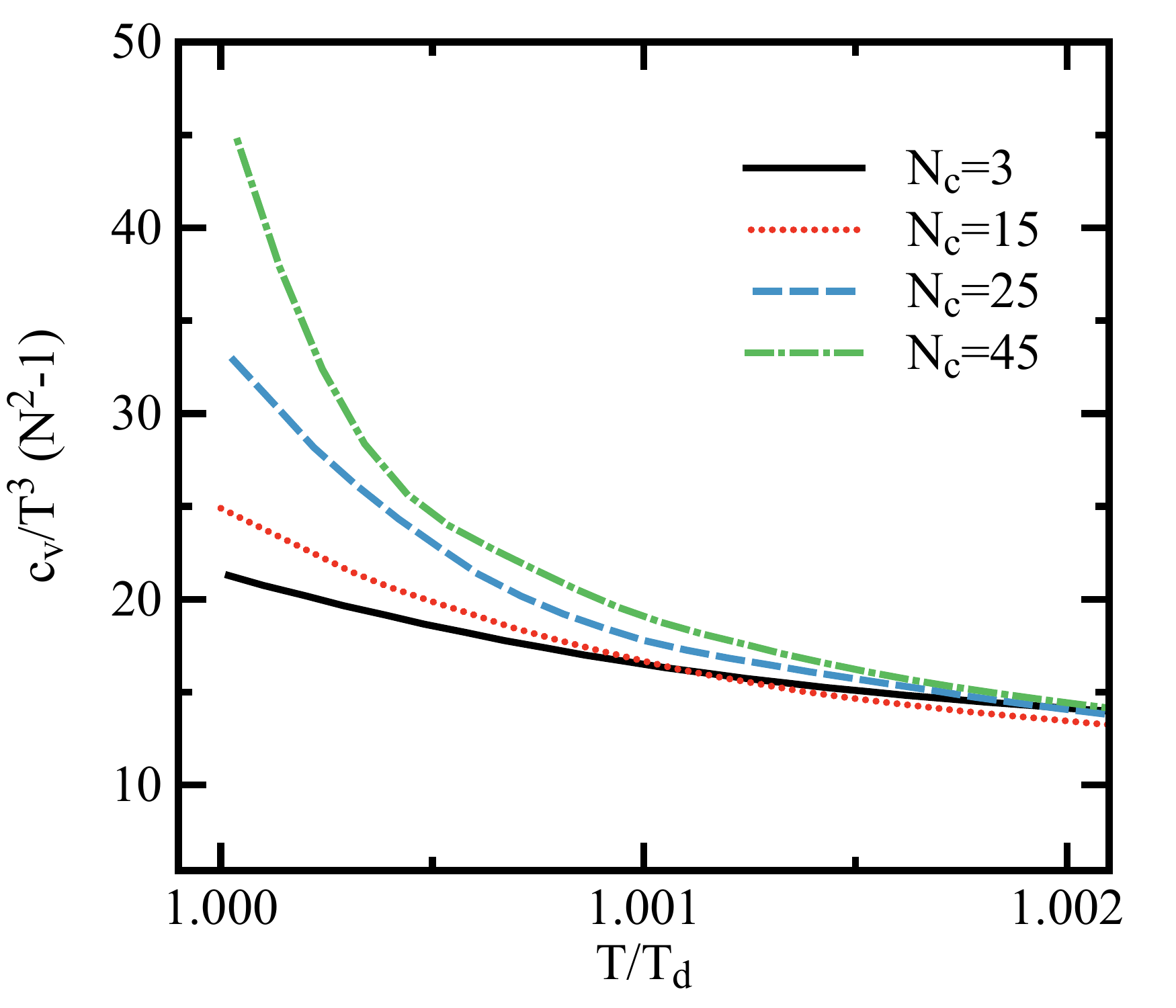} 
\caption{
Plot of the
specific heat, divided by $(N^2-1) T^3$, for different values of $N$.
}
\end{figure}

\begin{acknowledgments}
The research of R.D.P. and V.S. is supported
by the U.S. Department of Energy under contract \#DE-AC02-98CH10886.
\end{acknowledgments}


%

\end{document}